\documentclass[preprint,11pt,3p,twocolumn]{elsarticle}

\usepackage{graphicx,colordvi}
\usepackage{amssymb,amsmath,amsthm,amsfonts,amsxtra,dsfont,float}
\usepackage{tikz}
\usepackage{hyperref,url}
\hypersetup{colorlinks=true}
\usepackage[caption=false,font=footnotesize]{subfig}

\biboptions{numbers,sort&compress}

\newcommand{\parenths}[1]{\left( #1 \right)}

\newcommand{\ve}[1]{\boldsymbol{#1}}

\journal{ArXiv}

\begin{document}

\begin{frontmatter}



\title{On the choice of metric in gradient--based theories of brain function}


\author[addr1,addr2]{Simone Carlo Surace}
\address[addr1]{Department of Physiology, University of Bern, Switzerland}
\address[addr2]{Institute of Neuroinformatics and Neuroscience Center Zurich, University Z\"urich and ETH Z\"urich, Switzerland}

\author[addr1,addr2]{Jean-Pascal Pfister}
\author[addr3]{Wulfram Gerstner}
\author[addr3]{Johanni Brea}
\address[addr3]{School of Computer and Communication Sciences and Brain Mind Institute, School of Life Sciences, \'Ecole Polytechnique Federale de Lausanne, Switzerland}

\address{}

\begin{abstract}
The idea that the brain functions so as to minimize certain costs pervades theoretical neuroscience.
Since a cost function by itself does not predict how the brain finds its minima, additional assumptions about the optimization method need to be made to predict the dynamics of physiological quantities.
In this context, steepest descent (also called gradient descent) is often suggested as an algorithmic principle of optimization potentially implemented by the brain.
In practice, researchers often consider the vector of partial derivatives as the gradient.
However, the definition of the gradient and the notion of a steepest direction depend on the choice of a metric.
Since the choice of the metric involves a large number of degrees of freedom, the predictive power of models that are based on gradient descent must be called into question, unless there are strong constraints on the choice of the metric.
Here we provide a didactic review of the mathematics of gradient descent, illustrate common pitfalls of using gradient descent as a principle of brain function with examples from the literature and propose ways forward to constrain the metric.
\end{abstract}


\end{frontmatter}


\section{Introduction}
\label{}
The minimization of costs is a widespread approach in theoretical neuroscience \cite{Barlow1989,Bell1995,Olshausen1996,Dayan2005,Bialek2012}.
Cost functions that have been postulated range from energy consumption, free energy, negative entropy, reconstruction error, to distances between distributions that form representations of the world \cite{Barlow1989,Bell1995,Olshausen1996,Stemmler1999,Bohte2002,Booij2005,Bohte2005,Lengyel2005,Pfister2006,Gtig2006,Bialek2012,Galtier2013,Xu2013,Brea2013,Urbanczik2014,Buckley2017}.
In some cases, cost as performance of a biological system is measured in comparison to the absolute physical minimum \cite{Bialek2012} or an information theoretic optimum \cite{Barlow1989,Bell1995,Olshausen1996} without addressing the question of how a solution at or close to the minimum can be found.
In other cases, cost is used to derive algorithms that move the system closer to the minimum \citep{Stemmler1999,Bohte2002,Booij2005,Bohte2005,Lengyel2005,Pfister2006,Gtig2006,Triesch2007,Galtier2013,Xu2013,Brea2013,Urbanczik2014,Buckley2017,Costa:2017dk}.
In the second case, predictions entail update rules of neuronal quantities or differential equations for the time evolution of synaptic weights.

Optimization methods to train neural network models are often taken from machine learning, a field that has had intense interactions with theoretical and computational neuroscience \cite{Marblestone2016}.
A successful method in machine learning -- despite its simplicity -- has been the method of (stochastic) \emph{steepest descent} or \emph{gradient descent} \cite{Goodfellow2016}.

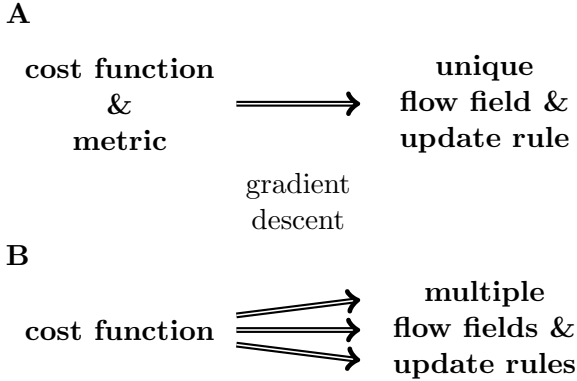
\begin{figure}[]
   \centering
   \begin{tikzpicture}[every node/.style = {anchor = west}]
        \node (a) at (0, .2) {\textbf{A}};
        \node (cfpm) at (.1, -1) {\parbox{2.8cm}{\centering\bf cost function\\\&\\metric}};
        \node (us) at (4.8, -1) {\parbox{3cm}{\centering\bf unique \\flow field \&\\ update rule}};
        \draw[thick, double, ->] (cfpm) -- node[midway, above=-1.8cm] {\parbox{2cm}{\centering gradient\\ descent}} (us);
        \node (a) at (0, -3) {\textbf{B}};
        \node (cfpm) at (.1, -4) {\parbox{2.8cm}{\centering\bf cost function}};
        \node (us) at (4.8, -4) {\parbox{3cm}{\centering\bf multiple \\flow fields \&\\ update rules}};
        \draw[thick, double, ->] (cfpm) -- (us);
        \draw[thick, double, ->] (cfpm) -- ([yshift=4mm]us.west);
        \draw[thick, double, ->] (cfpm) -- ([yshift=-4mm]us.west);
    \end{tikzpicture}
\caption{The main message of this text.
\textbf{A} A cost function and a metric together determine a unique flow field and update rule, given by gradient descent on the cost function in that metric.
\textbf{B} For a given cost function there are infinitely many different flow lines and update rules (one for each choice of the metric) that lead to the minima of the cost function by gradient descent.
}
\label{fig1}
\end{figure}

Gradient descent and steepest descent are the same, since the negative gradient points in the direction of steepest descent (see \autoref{eq:steepestascent}).
Often the direction of gradient descent is visualised as a vector orthogonal to the contour lines of the cost function.
The notion of orthogonality, however, assumes a \emph{Riemannian metric} (also known as inner product or scalar product in vector spaces).
The Riemannian metric enters also in an alternative, but equivalent definition of the direction of steepest descent:
The direction of steepest descent produces the greatest absolute decrease of the cost function for a step of a fixed (and small) size, where the step size is determined by the choice of the Riemannian metric.
Thus, a cost function by itself does not predict the trajectories that lead to its minima through steepest descent, however, a cost function combined with a metric does (see Figure~\ref{fig1}).

\begin{figure*}[ht]
    \centering
    \begin{tabular}{ll}
        \textbf{A} & \textbf{B}  \\
        \includegraphics{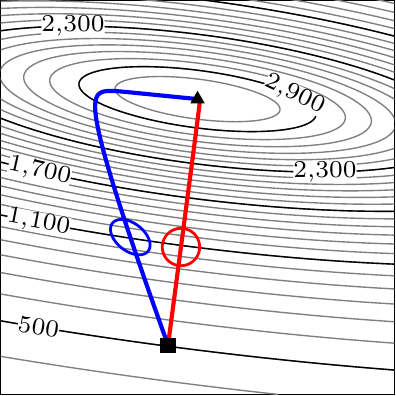} & \includegraphics{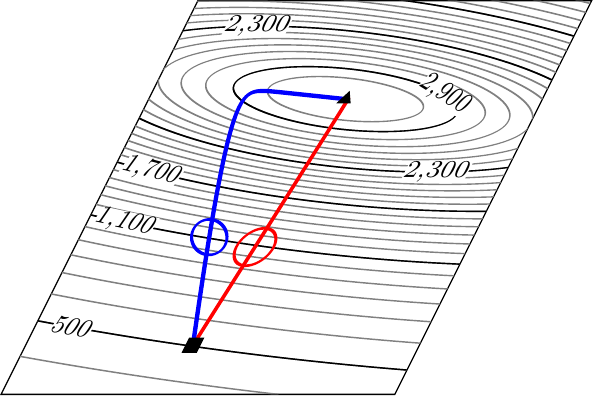}
    \end{tabular}
    \caption{\textbf{The ``wicked-map problem''}.
        \textbf{A} An ambitious mountaineer may follow the gradient in Euclidean metric to reach the mountain top (\textbf{red route} from square to triangle).
        Since the map is plotted in Cartesian coordinates, the route stands perpendicular to the contour lines.
        \textbf{B} If the ambitious mountaineer does not realise that a map given by a wicked hotelier is sheared, the \textbf{blue route} would be chosen, as it is now the one that stands perpendicular to the contour lines in the sheared map.
        The blue route corresponds to gradient ascent in another metric.
        Of course, each route on the normal map could be transformed to the sheared map and vice versa, but what looks like naive (Euclidean) gradient ascent in one map may look different in another map.
    }
    \label{fig:toyfig}
\end{figure*}

Why do we normally not think of the metric as an important and essential quantity?
The physical space that surrounds us, at the scales that we encounter in everyday life, is Euclidean.
Thus, a mountaineer who would like to determine the direction of steepest ascent of the terrain refers to Euclidean geometry.
In this case, the steepest direction is unambiguous because the way to measure distances is intrinsic to the space and not merely an artifact of using a particular set of coordinates.
On a map that faithfully represents Euclidean geometry, i.e. preserves angles and lengths up to some scaling factor, the mountaineer may find the steepest direction by drawing a curve that runs perpendicular to the contour lines (see \autoref{fig:toyfig}A red route).
But if a wicked hotelier gave the mountaineer a map that does not faithfully represent Euclidean geometry, another route would be chosen when planning the route as perpendicular to the contour lines (see \autoref{fig:toyfig}B blue route).
We will refer to this as the ``wicked-map problem'' in the following.

What may look obvious in the context of hiking maps can be confusing in contexts in which it is less clear how to draw a sensible map, i.e. how to choose a natural parametrization of an observed phenomenon.
We will discuss, how naive gradient ascent or descent, as taught in text books (e.g. \cite{Dayan2005,Goodfellow2016}), is susceptible to the ``wicked-map problem''.
While it is simple to display the same path in different maps by following standard transformation rules, the choice of an appropriate metric remains a challenge.
In other words, how should one know a priori which metric is most appropriate to predict a route with gradient ascent dynamics?
We will illustrate the problems around gradient ascent and descent with three examples from the theoretical neuroscience literature and discuss ways forward to constrain the choice of metric.

\section{The gradient is not equal to the vector of partial derivatives}\label{sec:problems}
Given a cost function $C(\ve x)$ that depends on variables $\ve x = (x_1,\ldots, x_N)$, where the variables $x_i$ could be synaptic weights or other plastic physiological quantities, naive gradient descent dynamics is sometimes written as \cite{Dayan2005,Goodfellow2016}
\begin{equation}
    x_{i} \rightarrow x_i - \tilde\eta \frac{\partial C(\ve x)}{\partial x_i}\, ,\label{eq:naivegraddesc}
\end{equation}
or in continuous time
\begin{equation}
    \frac{d}{dt}x_i(t) = -\eta\frac{\partial C(\ve x)}{\partial x_i} \label{eq:naivegraddesc2}
\end{equation}
where $\tilde\eta$ and $\eta$ are parameters called learning rate.
As we will illustrate in the course of this section, this has two consequences:
\begin{itemize}
    \item The ``wicked-map problem'': the dynamics in \autoref{eq:naivegraddesc} and \autoref{eq:naivegraddesc2} depend on the choice of the coordinate system.
    \item The ``unit problem'': if $x_i$ has different physical units than $x_j$, the global learning rate $\eta$ should be replaced by individual learning rates $\eta_i$ that account for the different physical units.
\end{itemize}
In Section~\ref{Sec3}, we will explain the geometric origin of these problems and how they can be solved.

The ``wicked-map problem'' often occurs in combination with the ``unit problem'', but it is present even for dimensionless parameters.
The parameters or coordinates that are used in a given problem are mostly arbitrary; they are simply labels attached to different points -- while the points themselves (for example, the position of the mountaineer) have properties independent of the parameters chosen to represent them.
For example, it is common to scale the variables or display a figure in logarithmic units, or simply display them in a different aspect ratio (transformations like the shearing transformation in \autoref{fig:toyfig}).
We expect the predictions of a theory to be independent of the choice of parametrizations.
Hence, if we think of the optimization as a biophysical process that effectively minimizes a cost function, then this biophysical process should not depend on our choice of the coordinate system.
However, as we will show below, a rule such as \autoref{eq:naivegraddesc2} that equates the time derivative of a coordinate with the partial derivative of a cost function (times a constant) is \emph{not} preserved under changes of parametrization (see \autoref{fig:toyfig}A,B).

In order to address the ``unit problem'' we can normalize each variable by dividing by its mean or maximum so as to make it unitless.
However, this merely replaces the choice of an arbitrary learning rate $\eta_i$ for each component by the choice of an arbitrary normalizing constant for each variable.

\subsection{Artificial examples}
To illustrate the ``wicked-map problem'', let us first consider the minimization of a (dimensionless) quadratic cost $C(x)=(x-1)^2$, where $x>0$ is a single dimensionless parameter.
The derivative of $C$ is given by $C'(x)=2x-2$.
Naive gradient descent minimization according to \autoref{eq:naivegraddesc2} yields $\eta^{-1} \frac{d}{dt} x(t)=-C'(x(t))=2-2x(t)$ with solution $x(t) = 1 + e^{-2\eta t}$ for initial condition $x(0) = 2$.

\begin{figure*}[!ht]
\centering
\begin{tabular}{lll}
    \textbf{A} & \textbf{B} & \textbf{C}\\
    \includegraphics{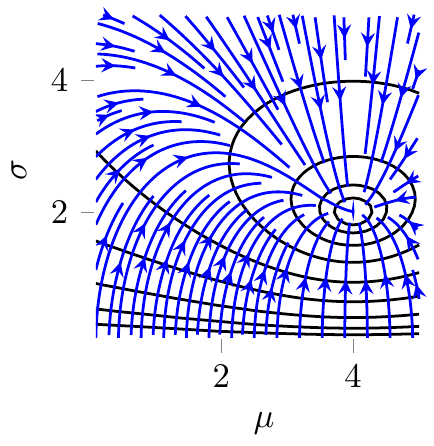} & \includegraphics{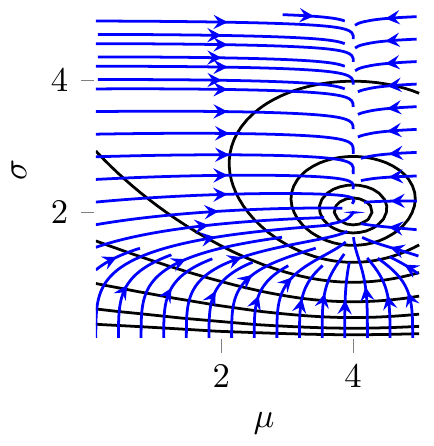} & \includegraphics{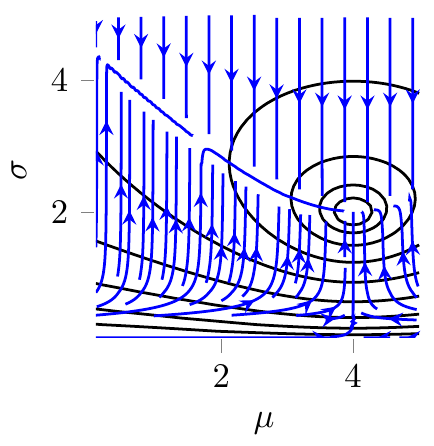} \\
    \includegraphics{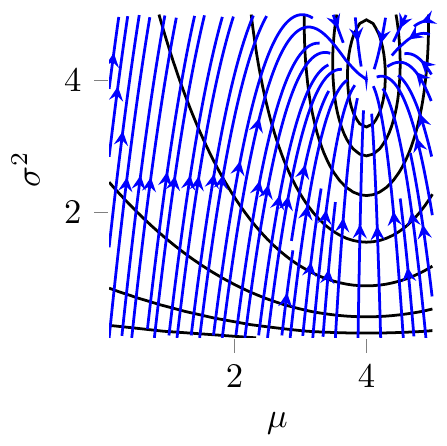} & \includegraphics{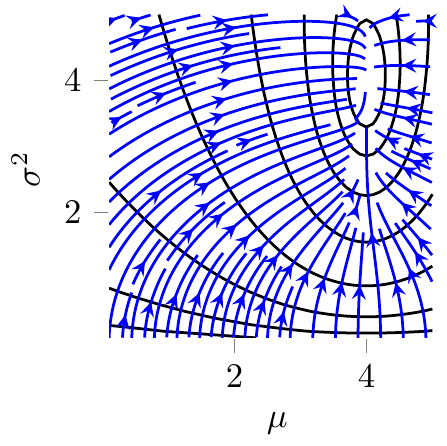} & \includegraphics{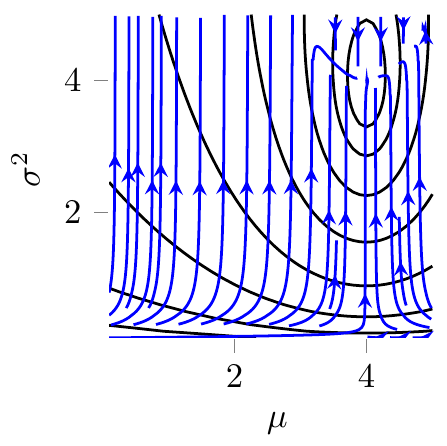}
\end{tabular}
\caption{\textbf{Minimizing the Kullback-Leibler divergence from a fixed normal distribution with mean 4 and standard deviation 2 to a parametrized normal distribution}.
Equipotential curves in black; flow fields generated by gradient descent in blue with \textbf{A} Euclidean metric in mean $\mu$ and standard-deviation $\sigma$, displayed in $\mu-\sigma$--plane (top row) and $\mu-\sigma^2$--plane (bottom row), \textbf{B} Euclidean metric in mean $\mu$ and variance $s = \sigma^2$, displayed in $\mu-\sigma$--plane (top row) and $\mu-\sigma^2$--plane (bottom row) and \textbf{C} Euclidean metric in mean $\mu$ and precision $\tau = 1/\sigma^2$, displayed in $\mu-\sigma$--plane (top row) and $\mu-\sigma^2$--plane (bottom row).}
\label{fig:kldescent}
\end{figure*}

Since $x$ is larger than zero and dimensionless, one may choose an alternative parametrization $\tilde x=\sqrt x$.
The cost function in the new parametrization reads $\tilde C(\tilde x)=(\tilde x^2 - 1)^2$, and its derivative is given by $\tilde C'(\tilde x)=4\tilde x(\tilde x^2 - 1)$.
In this parametrization, it may be argued that a reasonable optimization runs along the trajectory $\eta^{-1}\frac{d}{dt}{\tilde{x}}(t)=-\tilde C'(\tilde x(t))=-4\tilde x(\tilde x^2 - 1)$ with solution $\tilde x(t) = \frac1{\sqrt{-\frac12 e^{-8\eta t} + 1}}$ for initial condition $\tilde x(0) = \sqrt{2}$.
After transforming this solution back into the original coordinate system with parameter $x$, we see that the original dynamics $x(t) = 1 + e^{-2\eta t}$ and the new dynamics $\big(\tilde x(t)\big)^2 = \frac1{-\frac12 e^{-8\eta t} + 1}$ are very different.
This is expected, because the (1-dimensional) vector field $C'(x)=2-2x$ that is used for the first trajectory, should behave as $C'(x)\to \frac{\partial\tilde x}{\partial x} C'(x) = \frac1{\tilde x}-\tilde x$ under a change of parametrization, which is different from the vector field $\tilde C'(\tilde x)=-4\tilde x(\tilde x^2 - 1)$ that is used for the second trajectory.
This first, one-dimensional example shows that the naive gradient descent dynamics of \autoref{eq:naivegraddesc2} does not transform consistently under a change of coordinate system.

As a second example, consider the minimization by gradient descent of the cost function $C(\mu,\sigma)=D_{\text{KL}}(\mathcal{N}(\mu_0,\sigma_0)||\mathcal{N}(\mu,\sigma))$, the Kullback-Leibler divergence from a fixed normal distribution $\mathcal{N}(\mu_0,\sigma_0)$ to a normal distribution $\mathcal{N}(\mu,\sigma)$ parametrized by its mean $\mu$ and standard deviation $\sigma$.
A naive gradient descent dynamics would be given by 
$\tfrac{d\mu}{dt}=-\frac{\partial C}{\partial\mu}$ and $\tfrac{d\sigma}{dt}=-\frac{\partial C}{\partial\sigma}$.
The corresponding flow-field is shown in \autoref{fig:kldescent}A.

\begin{figure*}[!ht]
   \centering
   \begin{tabular}{ll}
       \textbf{A} & \textbf{B} \\
       \includegraphics{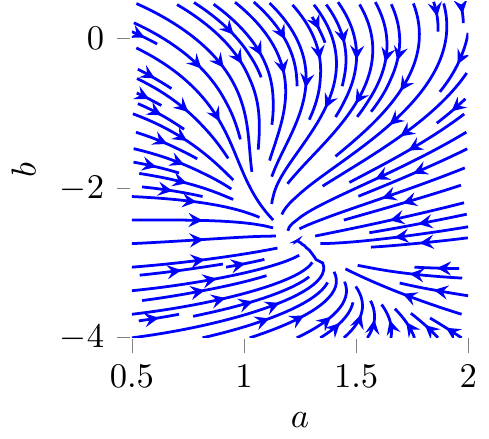} & \includegraphics{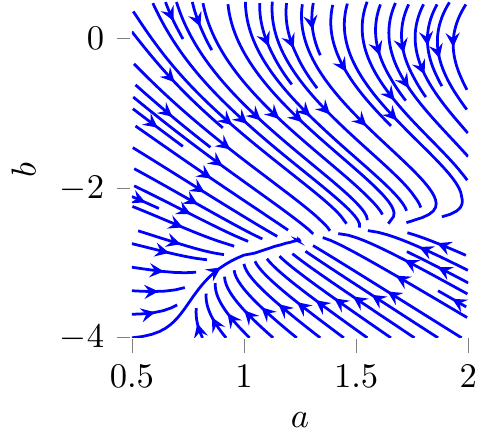}\\
       \textbf{C} & \textbf{D} \\
       \includegraphics{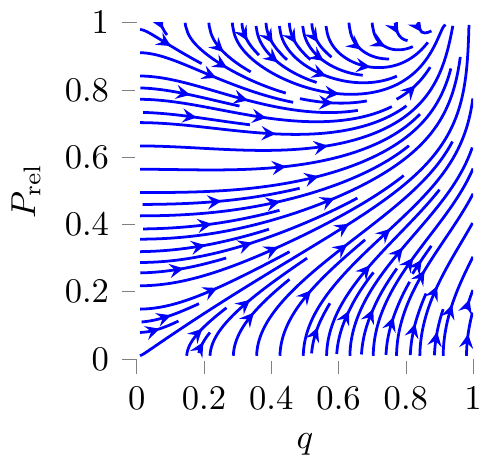} & \includegraphics{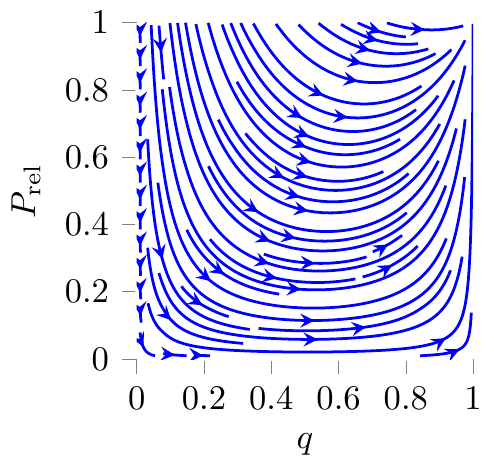}
   \end{tabular}
   \caption{\textbf{Gradient descent flow fields in neuroscience.}
       \textbf{A} Flow of intrinsic plasticity parameters $a$ and $b$ with Euclidean metric (see Figure 1A in \protect\cite{Triesch2007}) and \textbf{B} with Fisher information metric.
       \textbf{C} Flow  of quantal amplitude $q$ and release probability $P_\mathrm{rel}$ in a binomial release model of a synapse with Euclidean metric (see Figure 1D in \protect\cite{Costa:2017dk}) and \textbf{D} with Fisher information metric.
       Under other choices of the metric (see \autoref{sec:choosemetric}) the flow fields would again look different.
   }\label{fig:neuro}
\end{figure*}

Besides this parametrization, other equivalent ways to parametrize the normal distribution are mean $\mu$ and variance $s = \sigma^2$ or mean $\mu$ and precision $\tau = 1/\sigma^2$.
Thus the function $C$ is expressed in the other parametrizations
as $\tilde C(\mu,s) = C(\mu,\sqrt{s})$ or $\bar C(\mu,\tau)=C(\mu,1/\sqrt{\tau})$.
When we apply the same recipe as before to the new parametrizations, we obtain the dynamics $\frac{d\mu}{dt} = -\frac{\partial \tilde C}{\partial \mu}$ and $\frac{ds}{dt} = -\frac{\partial \tilde C}{\partial s}$ and similar expressions for $\bar C$.
The corresponding flow-fields in \autoref{fig:kldescent}B and C differ from the one obtained with the initial parametrization (\autoref{fig:kldescent}A) and from each other.

This can also be seen by applying the chain rule to the two sides of $\frac{ds}{dt} = -\frac{\partial \tilde C}{\partial s}$ and comparing the result to $\frac{d\sigma}{dt} = -\frac{\partial C}{\partial \sigma}$, the dynamics in the original parametrization.
On the left hand side we get $\frac{ds}{dt} = \frac{\partial s}{\partial \sigma}\frac{d\sigma}{dt}$, i.e. a pre-factor $\frac{\partial s}{\partial \sigma}$.
On the right hand side we get $-\frac{\partial \tilde C}{\partial s} =- \frac{\partial \sigma}{\partial s}\frac{\partial C}{\partial \sigma}$, i.e. a pre-factor $\frac{\partial \sigma}{\partial s}$.
If the dynamics in the new parametrization would be the same as the one in the initial parametrization, the two pre-factors would be the same.

Despite the different looks of the flow fields resulting from the three different parametrizations, all of them can be seen to describe dynamics that minimize the cost function (Figure~\ref{fig:kldescent}).
However, this example illustrates an important geometrical property that we will come back to later: the differential of a function $f$, i.e. the collection of its partial derivatives, does not transform like a proper vector.

\subsection{Gradient descent in neuroscience}
In this section we present three examples from published works, where it is postulated that the dynamics of a quantity relevant in neuroscience follows gradient descent on some cost function.

In 2007  a learning rule for intrinsic neuronal plasticity has been proposed to adjust two parameters $a,b$ of a neuronal transfer function $g_{ab}(x)=(1+\exp(-(ax+b)))^{-1}$ \cite{Triesch2007}.
The rule was derived by taking the derivatives of the Kullback-Leibler (KL) divergence $D_{\text{KL}}(f_y||f_{\text{exp}})$ between the output distribution $f_y$ resulting from a given input distribution over $x$ and the above transfer function, and an exponential distribution $f_{\text{exp}}$ with decay parameter $\mu>0$.
The flow field in Figure 1A of \cite{Triesch2007} (here \autoref{fig:neuro}A) is obtained with the Euclidean metric.
If $x$ is a current or a voltage, one would encounter the ``unit problem'', since $a$ and $b$ would have different physical units; one may therefore assume that $x$ is normalized such that $x$, $a$ and $b$ are dimensionless.
The ``wicked-map problem'' appears, since it is unclear whether the Euclidean distance in the $(a, b)$-plane is the most natural way to measure distances between the output distributions $f_y$ that are parametrized by $a$ and $b$.
In fact, in 2013 a different dynamics has been predicted for the same cost function, but under the assumption of the Fisher information metric\footnote{See Section~\ref{sec:choosemetric} for more details on the Fisher metric.}
 \cite{Neumann:2013ha} which can be considered a more natural choice to measure distances between distributions than the Euclidean metric (see \autoref{fig:neuro}B).

Similarly, it has been argued that the quantal amplitude $q$ and the release probability $P_{\text{rel}}$ in a binomial release model of a synapse  evolve according to a gradient descent on the KL divergence from an arbitrarily narrow Gaussian distribution
with fixed mean $\varphi$ to the Gaussian approximation of the binomial release model \cite{Costa:2017dk}.
To avoid the ``unit problem'', the quantal amplitude $q$ was appropriately normalised.
Since $q$ and $P_\text{rel}$ parametrize probability distributions, one may also argue for this study, that the Fisher information metric (\autoref{fig:neuro}D) is a more natural choice, a priori, than the Euclidean metric (\autoref{fig:neuro}C), but the corresponding flow fields are just two examples of the infinitely many possible flow fields that would be consistent with gradient descent on the same cost function.
Alternatively, one could e.g. consider metrics that depend on metabolic costs; it may be more costly to move a synapse from release probability $P_\text{rel}=0.9$ to release probability $P_\text{rel}=1.0$ than from $P_\text{rel}=0.5$ to $P_\text{rel}=0.6$.
If there is no further principle to constrain the choice of metric (see e.g. \autoref{sec:choosemetric}), data itself may guide the choice of metric.
Surprisingly, the available and appropriately normalized experimental data is consistent with the Euclidean metric in $P_\mathrm{rel} - q$ space \cite{Costa:2017dk}, but there is probably not sufficient data to discard a metric based on metabolic cost.

Gradient descent has been popular as an approach to postulate synaptic plasticity rules \cite{Bohte2002,Booij2005,Bohte2005,Pfister2006,Gtig2006,Galtier2013,Xu2013,Brea2013,Urbanczik2014,Buckley2017}.
As an example, minimizing by gradient descent the KL divergence from a target distribution of spike trains to a model distribution of spike trains \cite{Brea2013} is claimed to lead to the plasticity rule with a constant learning rate $\eta$.
This choice of a constant learning rate is equivalent to choosing the Euclidean metric on the weight space.
But there is no reason to assume that the learning rate should be constant or the same for each parameter (synaptic weight): one could just as well choose individual learning rates $\eta_{ij}(w_{ij})$.
This generalization corresponds still to the choice of a diagonal Riemannian metric.
But, while it is often assumed that the change of a synapse depends only on pre- and postsynaptic quantities (but see \cite{Brea2013}), it could be that there is some cross-talk between neighbouring synapses, which could be captured by non-diagonal Riemannian metrics.
This example shows that gradient descent does not lead to unique learning rules.
Rather, each postulate of a gradient descent rule should be seen as a family of possibilities: there is a different learning rule for each choice of the Riemannian metric.

\section{What is the gradient, then? How to do steepest descent in a generic parameter space.}
\label{Sec3}
In the preceding section, we have shown that the partial derivatives with respect to the parameters do not transform correctly under changes of parametrization (i.e. not as we would expect for the components of a vector or flow field).
In order to work with generic spaces which may carry different parametrizations, it is useful to apply methods from differential geometry.

A Riemannian metric on an $N$-dimensional manifold (an intrinsic property of the space) gives rise to an inner product (possibly position-dependent) on $\mathbb R^N$ for each choice of parametrization.
The matrix representation of the inner product depends on the choice of parametrization.
However, the dependence is such that the result of an evaluation of the inner product is independent of the choice of parametrization.
When described in this language, the geometry of the trajectories in the space is therefore independent of parameter choices.

We refer to the Appendix for a detailed treatment of the gradient in Riemannian geometry.
In the following, we simply give the definition of the gradient in terms of the inner product.

For a function $f:\mathbb R^N\rightarrow \mathbb R$ and an inner product $\langle \cdot, \cdot \rangle:\mathbb{R}^N\times\mathbb{R}^N\rightarrow \mathbb R$, a common implicit definition (e.g. \cite{Rudin1976}) of the gradient $(\nabla f)(\ve x)$ of $f$ at point $\ve x$ is
\begin{equation}
    \langle (\nabla f)(\ve x), \ve u\rangle = \lim_{\epsilon\to 0}\frac{f(\ve x
    + \epsilon \ve u) - f(\ve x)}{\epsilon}\, ,\label{eq:defgrad}
\end{equation}
for all non-zero vectors $\ve u\neq \ve 0$, i.e. the gradient $(\nabla f)(\ve x)$ is the vector that is uniquely defined by the property that its product with any vector $\ve u$ is equal to the derivative of $f$ in direction $\ve u$.
With the Euclidean inner product $\langle \ve v, \ve w\rangle_\mathrm{E} = \sum_{i=1}^N v_i w_i$ it is a simple exercise to see that the components of the gradient are the partial derivatives.
However, with any other inner product $\langle \ve v, \ve w\rangle_{G(\ve x)} = \sum_{i,j=1}^N v_i G_{ij}(\ve x) w_i$, characterized by the position-dependent symmetric, positive definite matrix $G(\ve x)$, the gradient is given by
\begin{equation}
    ( \nabla f)(\ve x) = G^{-1}(\ve x) \left(\begin{array}{c}\frac{\partial f}{\partial
            x_1}\\\vdots\\\frac{\partial f}{\partial
x_N}\end{array}\right)\, ,\label{eq:generalgrad}
\end{equation}
i.e. the matrix product of the inverse of $G(\ve x)$ with the vector of partial derivatives.
Note that the inverse $G^{-1}(\ve x)$ is also a symmetric, positive definite matrix.
The inverse of $G(\ve x)$ automatically carries the correct physical units and the correct transformation behaviour under reparametrizations, i.e. the components of the matrix transform as $G'_{ij} = \sum_{kl}\frac{\partial x_k}{\partial x'_i}\frac{\partial x_l}{\partial x'_j}G_{kl}$ under a reparametrization from $\ve x$ to $\ve x'$, such that the dynamics
\begin{equation}
    \frac{d}{dt} \ve x(t)= -\eta ( \nabla f)(\ve x(t))
    \label{eq:graddescent}
\end{equation}
is invariant under a change of parametrization.
Following standard nomenclature, we call the gradient induced by the Riemannian metric $G$ the \emph{Riemannian gradient}.

The gradient is used in optimization procedures because it points in the direction of steepest ascent. To see this we define the direction of steepest ascent
\newcommand{\argmax}[1]{\underset{#1}{\operatorname{argmax}}}
\begin{equation}
    \ve s(\ve x) \doteq \argmax{\langle \ve u, \ve u\rangle = 1}\lim_{\epsilon\to 0}\frac{f(\ve x
        + \epsilon \ve u) - f(\ve x)}{\epsilon} \\
\end{equation}
as the direction $\ve u$ in which the change of the function $f$ is maximal.
Using the definition of the gradient in \autoref{eq:defgrad} and determining the maximum we find
\begin{equation}
\begin{split}
     \ve s(\ve x)   &= \argmax{\langle \ve u, \ve
    u\rangle = 1} \langle(\nabla f)(\ve x), \ve u\rangle \\
    &= \frac{(\nabla f)(\ve
    x)}{||(\nabla f)(\ve
    x)||}\, ,\label{eq:steepestascent}
\end{split}
\end{equation}
where $||\cdot||=\sqrt{\langle\cdot,\cdot\rangle}$ denotes the  norm induced by the metric $\langle \cdot, \cdot \rangle$.

\section{On choosing a metric}\label{sec:choosemetric}
Given an arbitrary vector field one may ask whether it is possible to represent it as a steepest descent on some cost function with respect to some metric.
When the metric is already known there is a systematic way to check whether the vector field can be written as a gradient, and to construct a suitable cost function (see \ref{AppC}).
If the metric is unknown, one may have to construct a metric which is tailored to the dynamical system (see also \ref{AppC}).

Instead of constructing a custom-made metric for the dynamical system, it may be more desirable (from the perspective of finding the most parsimonious description) to choose a metric \emph{a priori} and then check whether a given dynamical system has the form of a gradient descent with respect to that metric.
Such an \emph{a priori} choice could be guided e.g. by biophysical principles and therefore becomes an integral part of the theory.
For example, a metric could reflect the equivalence of metabolic cost that is incurred in changing individual parameters.
Another example is Weber's law, which implies that parameter changes of the same \emph{relative} size are equivalent.
This would suggest a constant (but not necessarily Euclidean) metric on a logarithmic scale.
A third example is the homogeneity across an ensemble: if there are $N$ neurons of the same type and functional relevance, we may want to constrain the metrics to those that treat all neurons identically when changing quantities such as neuronal firing thresholds or synaptic weights.

Even if it does not fully determine the metric, a principle which constrains the class of metrics is very useful when trying to fit the metric to the data (i.e. for a given cost function).
Without any constraints, the specification of a Riemannian metric for an $n$-dimensional parameter space requires the specification of $\tfrac{1}{2}n(n+1)$ smooth functions, i.e. the components of the matrix $G$ in some coordinate system; these components can be constant or position-dependent.

If the parameter space describes a smooth family of probability distributions, the Fisher information matrix provides a canonical Riemannian metric on this manifold.
The special status of the Fisher-Rao metric in statistics is due to the fact that it is the only metric (up to scaling factors) that has a natural behavior under sufficient statistics (see e.g. \cite{Amari:2007ua}, Theorem 2.6 going back to Chentsov, 1972).
The Riemannian gradient with respect to the Fisher-Rao metric is often called the \emph{natural gradient}\footnote{Due to Chentsov's theorem, the Fisher information metric is regarded as a natural choice, but some authors (including Amari in \cite{Amari:2007ua}) seem to use the term \emph{natural gradient} to more broadly refer to a Riemannian gradient with respect to some metric that obeys some invariance principle.}, and has been applied in machine learning \cite{Amari:1998wc,Yang:1998wg,Kakade02,Peters06,Hoffman10,Desjardins:2015uo,Ollivier:2015cb,Ollivier:2015jz,Ollivier:2017wp} and neuroscience \cite{Neumann:2013ha}.
Another metric on probability distributions that has recently gained a lot of attention is the optimal transport or Wasserstein metric \cite{Ambrogioni:2018tw,Bernton:2017us,Arjovsky:2017vh}.
However, despite the nice mathematical properties of such metrics and their usefulness for machine learning applications, it is not clear why natural selection would favor them.
Therefore, the special mathematical status of those metrics does not automatically carry over to biology or more specifically neuroscience.

\section{Conclusions}
Steepest descent or gradient descent depends on a choice of ruler (or Riemannian metric) for the parameter space of interest.
The Euclidean metric is rarely a natural choice (see ``wicked-map problem'' in \autoref{sec:problems}), especially (but not only) for spaces of parameters that carry different physical units (see ``unit problem'' in \autoref{sec:problems}).
In practice, the ``unit problem'' can be treated with a suitable normalization of the measured quantities \cite{Triesch2007,Costa:2017dk}.
The ``wicked-map problem'', however, remains and it may be a matter of serendipity to select the parametrization in which naive gradient descent is consistent with experimental data.
Also when steepest descent is invoked to postulate the dynamics of firing rates or synaptic weights \cite{Bohte2002,Booij2005,Bohte2005,Pfister2006,Gtig2006,Galtier2013,Xu2013,Brea2013,Urbanczik2014,Buckley2017}, one should not ignore the possibility of non-Euclidean metrics.
The additional free hyperparameters associated with the choice of Riemannian metric can significantly alter the prediction of the model when those are about \emph{trajectories} along which optimization occurs (as opposed to just the targets of the optimization).
Unless there is an obvious way to fix those hyperparameters, be it through previously collected data or some principles, the model's predictive power is lowered, since many flow fields can be written as a gradient descent in some metric.
Whether a gradient descent model with many degrees of freedom or a phenomenological model without reference to the computational principle of gradient descent is preferable in this case, may be a matter of subjective preferences.

On the positive side, the additional degrees of freedom that accompany the choice of metric imply that a larger class of dynamics can be seen as being optimal in the sense of following a flow field consistent with gradient descent and some metric that is yet to be determined.
It will be interesting to uncover the metrics that are chosen by biology, and to uncover the biophysical principles that underlie these choices.
For it is the metric together with the cost function that fully specifies a gradient descent dynamics.

\section*{Acknowledgements}
We thank Rui Ponte Costa, Tim Vogels and Jochen Triesch for helpful feedback and Henning Sprekeler for the suggestion to use biophysical principles such as metabolic cost in order to constrain the metric.

\appendix
\section{Calculations for artificial example no. 2 (Fig. 2)}
The cost function is given by
\begin{equation}
\begin{split}
f(\mu,\sigma)&=\frac{(\mu-\mu_0)^2}{2\sigma^2}
\\&\quad+\log\frac{\sigma}{\sigma_0}+\frac{\sigma_0^2}{2\sigma^2}-\frac{1}{2},
\end{split}
\end{equation}
and its partial derivatives read
\begin{align}
\frac{\partial f}{\partial \mu} &= \frac{\mu_0-\mu}{\sigma^2}\\
\frac{\partial f}{\partial \sigma} &= \frac{(\mu_0-\mu)^2+\sigma_0^2-\sigma^2}{\sigma^3}.
\end{align}
The other partial derivatives can be computed either by using the chain rule, e.g.
\begin{equation}
\begin{split}
\frac{\partial\tilde f}{\partial s}(\mu,s)&=\frac{\partial\sigma}{\partial s}(s)\frac{\partial f}{\partial \sigma}(\mu,\sqrt{s})\\
&=\frac{1}{2\sqrt{s}}\times\frac{(\mu_0-\mu)^2+\sigma_0^2-s}{s^{3/2}} \\
&=\frac{(\mu_0-\mu)^2+\sigma_0^2-s}{2s^{2}},
\end{split}
\end{equation}
or by expressing the function in terms of the new coordinates, e.g.
\begin{equation}
\begin{split}
\bar f(\mu,\tau)&=f(\mu,1/\sqrt{\tau}), \\
&=\frac{1}{2}(\mu-\mu_0)^2\tau
\\&\quad-\log\sigma_0\sqrt{\tau}+\frac{\sigma_0^2\tau}{2}-\frac{1}{2}
\end{split}
\end{equation}
and then calculating the derivative directly:
\begin{equation}
\begin{split}
\frac{\partial\bar f(\mu,\tau)}{\partial\tau}&=\frac{1}{2}(\mu-\mu_0)^2+\frac{\sigma_0^2}{2}-\frac{1}{2\tau}.
\end{split}
\end{equation}
To define the corresponding vector fields, we follow the convention to denote the tangent vector in the direction of a parameter $\theta$ by $\partial_{\theta}$.
The vector fields $V,\tilde V$, and $\bar V$, defined by
\begin{align}
-V         	&=\tfrac{\partial f}{\partial\mu}\partial_{\mu}+\tfrac{\partial f}{\partial\sigma}\partial_{\sigma},\\
-\tilde V 	&=\tfrac{\partial\tilde f}{\partial\mu}\partial_{\mu}+\tfrac{\partial\tilde f}{\partial s}\partial_{s}, \\
-\bar V 	&=\tfrac{\partial\bar f}{\partial\mu}\partial_{\mu} +\tfrac{\partial\bar f}{\partial\tau}\partial_{\tau},
\end{align}
therefore read
\begin{align}
-V         	&=\tfrac{\mu_0-\mu}{\sigma^2}\partial_{\mu}+\tfrac{(\mu_0-\mu)^2+\sigma_0^2-\sigma^2}{\sigma^3}\partial_{\sigma},\\
-\tilde V 	&=\tfrac{\mu_0-\mu}{s}\partial_{\mu}+\tfrac{(\mu_0-\mu)^2+\sigma_0^2-s}{2s^{2}}\partial_{s}, \\
-\bar V 	&=\tfrac{\mu_0-\mu}{s}\partial_{\mu} +\parenths{\tfrac{(\mu-\mu_0)^2}{2} +\tfrac{\sigma_0^2}{2}-\tfrac{1}{2\tau}}\partial_{\tau}.
\end{align}
Let us express the fields $V,\tilde V$ in the other parametrizations that are displayed in \autoref{fig:kldescent}A,B.
The basis vectors $\partial_{\sigma}$ and $\partial_s$ are related by
\begin{equation}
\begin{split}
\partial_{\sigma}=\frac{\partial s}{\partial\sigma}\partial_s=2\sigma\partial_s=2\sqrt{s}\partial_s,
\end{split}
\end{equation}
which implies that $V$ can be expressed in $\mu,s$ coordinates as
\begin{align}
-V         	&=\tfrac{\mu_0-\mu}{\sigma^2}\partial_{\mu}+2\tfrac{(\mu_0-\mu)^2+\sigma_0^2-s}{s}\partial_{s},
\end{align}
and $\tilde V$ in $\mu,\sigma$ coordinates as
\begin{align}
-\tilde V    	&=\tfrac{\mu_0-\mu}{\sigma^2}\partial_{\mu}+\tfrac{(\mu_0-\mu)^2+\sigma_0^2-\sigma^2}{4\sigma^{5}}\partial_{\sigma}.
\end{align}
This shows that $V$ and $\tilde V$ are different.
We leave the corresponding calculation for the $\mu,\tau$ parametrization as an exercise.

\section{Steepest descent on manifolds}
\label{AppB}
Here, we give a short introduction on the \emph{calculus on manifolds} and \emph{differential geometry} background of this paper.
For more details, the reader is referred to the excellent books by Michael Spivak \cite{Spivak:1971ut} and by Jeffrey M. Lee \cite{Lee:2009vl}.

Let $p$ be a point in the manifold and $v$ be a vector from the tangent space in $p$.
Suppose that we want to define the directional derivative of $f$ at point $p$ in the direction $v$.
We can then draw a curve $\gamma$ that runs through $p$ and which has a tangent vector equal to $v$ at that point.
For convenience, let $\gamma(t)=p$.
We then define the differential $df_p$ of $f$ at $p$ as
\begin{equation}
df_p(v)=\frac{d}{dt}f\parenths{\gamma(t)}|_{t=0},
\end{equation}
i.e. as a map from the tangent space to the real numbers.
It can be shown that this map is linear and well-defined (i.e. it does not depend on the particular choice of $\gamma$).
In a parametrization $p=\Phi(x)=\Phi(x^1,...,x^n)$ it reads
\begin{equation}
df_p(v)=\sum_{i=1}^nv^i \partial_{i}f(\Phi(x)),
\end{equation}
where $\partial_{i}$ denotes the partial derivative with respect to $x^i$ and $v^{i}$ is the $i$'th component of $v$ when expressed in the coordinate basis.
Here, we introduced upper indices for tangent vectors and lower indices for so-called cotangent vectors.

As a linear map from the tangent space to the real numbers, $df_p$ belongs to the cotangent space at $p$, the dual space of the tangent space.
The cotangent vector $df_p$ is expressed as
\begin{equation}
df_p=\sum_{i=1}^n\partial_{i}f(\Phi(x)) dx^i
\end{equation}
in local coordinates.
Recall from linear algebra that the dual space $V^{\ast}$ of a real finite-dimensional vector space $V$ is isomorphic to $V$ but not in a canonical way, i.e. there is no preferred way to associate tangent vectors and cotangent vectors one-to-one.
Given a basis of $V$, one can choose a dual basis of $V^{\ast}$ and this gives rise to an isomorphism, whose representation in this basis is the unit matrix.
The same concept holds for the tangent and cotangent spaces of a smooth manifold.
A choice of coordinates $x_i$, $i=1,...,n$ gives rise to a basis of the tangent space $\partial_i$, $i=1,...,n$ and a corresponding dual basis $dx^i$, $i=1,...,n$.
Therefore the identification of tangent and cotangent vectors depends on the choice of coordinates.
It is for this reason that tangent and cotangent vectors have to be regarded as different objects.

The different geometrical nature of tangent and cotangent vectors is the fundamental reason why a rule such as in \autoref{eq:naivegraddesc} or \autoref{eq:naivegraddesc2} is problematic:
on one side of the equation, we have a tangent vector (the velocity vector of the curve along which we want to move), while on the other side we have the differential of the cost function, a cotangent vector.
They cannot be equal; they can at most have the same components in \emph{some} coordinates, but this property is lost when changing to a different set of coordinates.
Such a rule therefore does not make sense without invoking a preferential choice of coordinates.

\subsubsection*{Generalization of inner products: Riemannian metrics}
In order to obtain a way to transform cotangent vectors into tangent vectors or vice versa and thereby identify them with each other, one needs to define additional structure on the manifold.
This structure comes in the shape of what is called a Riemannian metric, which is a map from bivectors (i.e. pairs of tangent vectors) to the real numbers.
More specifically, at each point $p$ it specifies a quadratic form or an inner product $g_p$ on the tangent space at that point.
In order to qualify for the term \emph{Riemannian}, this quadratic form should in addition be positive definite.\footnote{In many contexts, e.g. in physics, however, metrics are pseudo-Riemannian.}
Lastly, the metric is usually expected to vary smoothly as a function of the position in the manifold, which means that when it is evaluated on smooth vector fields, the resulting real-valued function is smooth.
Given a Riemannian metric $g$ and a point $p$, a cotangent vector $v^{\flat}$ is assigned to a tangent vector $v$ in the following way
\begin{equation}
v^{\flat}: \quad v' \mapsto g_p(v,v'),
\end{equation}
or in local coordinates
\begin{equation}
v^{\flat}_i=\sum_{j=1}^n g_{ij} v^j, \quad v^{\flat}(v')=\sum_{i,j=1}^n g_{ij} v^i v'^j.
\end{equation}
Since $g_p$ is a bilinear form, we see that both $v^{\flat}$ itself (as a map from the tangent space to the reals) and the assignment of $v^{\flat}$ to $v$ are linear maps, and we can also see that the assignment is injective, because if it were otherwise, we could have
\begin{equation}
0=v_1^{\flat}(v')-v_2^{\flat}(v')=g_p(v_1-v_2,v').
\end{equation}
for non-zero $v_1\neq v_2$ and some non-zero $v'$, which contradicts the positive definiteness of $g_p$.
Since the tangent space and the cotangent space have the same dimension, the assignment is also surjective, and we can therefore define an inverse $^{\sharp}$ that assigns a tangent vector $\omega^{\sharp}$ to any cotangent vector $\omega$.
An inverse metric $g_p^{-1}$ may then be defined as
\begin{equation}
g^{-1}(\omega_1,\omega_2)=g_p(\omega_1^{\sharp},\omega_2^{\sharp}).
\end{equation}
In local coordinates, we may write
\begin{multline}
g^{-1}(\omega_1,\omega_2)=\sum_{i,j=1}^ng^{ij}\omega_{1,i}\omega_{2,j}, \\ \sum_{j=1}^ng_{ij}g^{jk}=\delta_j^k,
\end{multline}
where $\delta_j^k$ are the components of the unit matrix (i.e. $\delta_j^k=1$ if $j=k$ and zero otherwise).
Using this inverse metric, $\omega^{\sharp}$ may be written as
\begin{equation}
\omega^{\sharp}=g^{-1}(\omega,\cdot), \quad \omega^{\sharp,i}=\sum_{j=1}^n g^{ij}\omega_j.
\end{equation} Indeed, as a linear map from the cotangent space to the reals, the RHS may be canonically identified with a tangent vector.\footnote{For a finite-dimensional smooth manifold, the map $v\mapsto (\omega\mapsto\omega(v))$ is an isomorphism between the tangent space and its double dual.} The isomorphisms $^{\sharp}$ and $^{\flat}$ are known as \emph{musical isomorphisms}, and in terms of local coordinates, they are used to raise and lower indices.

In analogy to the case in $\mathbb R^N$ we can now define the gradient on smooth manifolds
\begin{equation}
    g_p(\nabla_p f, v) = df_p(v)\, ,
\end{equation}
for all tangent vectors $v$ at point $p$ and, using the definition of the inverse metric, we find
\begin{equation}
    \nabla_p f = df_p^\sharp\, .\label{eq:generalgradmanifold}
\end{equation}

\subsubsection*{The gradient on a Riemannian manifold}
By being given the structure of the Riemannian metric, we obtain a notion of lengths of and angles between tangent vectors, as with any other inner product space.
Thus, given a point $p$, we can ask in which direction the steepest ascent of the function $f$ is.
The answer is given by
\begin{equation}
s(p)\doteq\underset{g(v,v)=1}{\operatorname{argmax}}\;df_p(v),
\end{equation}
where the maximum is taken over all unit-length tangent vectors, and the directional derivative is properly expressed via the action of the differential of $f$ on the tangent vector.
This constrained optimization has a cost function
\begin{equation}
\mathcal{L}=df_p(v)-\lambda(g_p(v,v)-1),
\end{equation}
where $\lambda$ is a Lagrange multiplier.
In order to solve this optimization problem, we have to compute the differential of $\mathcal{L}$ with respect to $v$ and set it to zero.
Because of the linearity of $df_p$ and the symmetry and bilinearity of $g_p$, we have
\begin{equation}
\begin{split}
\mathcal{L}(v+v')&=df_p(v+v')\\
&\quad-\lambda(g_p(v+v',v+v')-1)\\
&=\mathcal{L}(v)+df_p(v') -2\lambda g_p(v,v')\\
&\quad-\lambda g_p(v',v').
\end{split}
\end{equation}
The critical tangent vector $v$ is therefore characterized by the vanishing of the term that is linear in $v'$
\begin{equation}
d\mathcal{L}(v')=df_p(v')-2\lambda g(v,v')=0,
\end{equation}
to be satisfied by all tangent vectors $v'$.\footnote{Note that the tangent space of the tangent space is the tangent space itself.} As we developed above, this equation has a unique solution, given by
$
v=\frac{1}{2\lambda}df_p^{\sharp},
$
which, when normalized, reads
\begin{equation}
v=\frac{df_p^{\sharp}}{\sqrt{g_p(df_p^{\sharp},df_p^{\sharp})}}\, ,
\end{equation}
which points in the same direction as the gradient in
\autoref{eq:generalgradmanifold}.

\section{Which dynamical systems can be regarded as a gradient descent on a cost function?}
\label{AppC}
In some cases we may start with a given dynamical system in the form of a vector field $V$ on some manifold $M$.
The question arises whether we can find a function $f$ and a metric $g$ such that the dynamical system takes the form of a (negative) gradient flow, i.e. $V=-\nabla_g f$.
For this question to make sense, we fix an asymptotically stable set $S$, with domain of attraction $A$.

If $g$ is given, e.g. from the considerations of the previous section, but we do not know $f$, we can compute the one-form $V^{\flat}$ that is dual to $V$ with respect to $g$, and check whether it is closed.\footnote{In three-dimensional space, this reduces to checking whether $\text{curl }V=0$.}
If $V^{\flat}$ is indeed closed and the domain of attraction $A$ is contractible (this is always true if $S$ consists of a single point), this implies the existence of a function $f$, unique up to an additive constant, such that $V^{\flat}=-df$, and hence $V=-\nabla_g f$, on $A$.
A suitable potential function $f$ may be found by picking a reference point $p_0\in S$ and integrating $V^{\flat}$ along any curve that joins $p_0$ and $p$.
Note that if we change to a different metric $g'$, the corresponding $V^{\flat}$ might no longer be closed and hence such a potential may cease to exist.

If neither $f$ nor $g$ are given, necessary and sufficient conditions for their existence on a compact manifold were given by \cite{Smale:1961dj,Wall:1972gv} in terms of transversality conditions on the vector field: it needs to be transversal to the zero section at each fixed point, transversal to the boundary, and the stable and unstable manifolds of each fixed point have to meet transversally.
On a non-compact manifold it is not known whether we can find a global metric, but we can always use the construction above on a compact subset.
Alternatively one may find a smooth Lyapunov function (this is always possible; see Theorem 3.2 in \cite{WilsonJr:1969ul}) and use the method in the next paragraph to construct a suitable metric.

Suppose that $f$ is given, and $g$ is sought.
This case is discussed in \cite{Wall:1971bk}.
A necessary condition for the existence of $g$ is that $f$ is a smooth local Lyapunov function for $V$, i.e. $f>0$ and $Vf=df(V)<0$ on $A\backslash S$, and $f=0$ on $S$.
But this may not be sufficient:
a simple counterexample is the dynamical system $dx/dt=V(x)=-x$ and $f(x)=x^4$ on $\mathbb{R}$.
This dynamical system has a global attractor at $x=0$, and $f$ is a global Lyapunov function since we have $f(0)=0$ as well as $f(x)>0$ and $df(V)=4x^3(-x)=-4x^4<0$ for all $x\neq 0$.
But if we want to write $dx/dt=-df(x)/g(x)$, we obtain $g(x)=-df(x)/V(x)=4x^2$, which is not a Riemannian metric on $\mathbb{R}$ (not positive definite at $x=0$).

If we are happy to exclude the set $S$, we can always find a Riemannian metric defined on $A\backslash S$ such that $V$ is the negative gradient of a given smooth local Lyapunov function $f$:
in the one-dimensional example above, we just have to divide the negative differential of $f$ by $V$.
In higher dimensions, we may consider the level sets $f^{-1}(q)$ for $q\in(0,a)=f(A\backslash S)$, which are submanifolds of dimension $n-1$.
We may then choose a Riemannian metric on each level set such that it depends smoothly on $q$, and extend this to a Riemannian metric on $M$ by declaring $V$ to be orthogonal to the level sets and having a squared Riemannian length equal to $|df(V)|$.
The conditions for being able to extend this to a metric on $A$ are discussed in \cite{Wall:1971bk}.

\bibliographystyle{elsarticle-num}
\bibliography{bibliography}

\end{document}